\documentclass[prl,twocolumn,nofootinbib, preprintnumbers, superscriptaddress]{revtex4}

\usepackage{slashed}
\usepackage{amsmath,amssymb,slashed,braket}
\usepackage{graphicx}
\usepackage{epstopdf}
\usepackage{float,appendix}
\usepackage[colorlinks=true,
            linkcolor=blue,
            urlcolor=blue,
            citecolor=green,          
            bookmarks=true,
            bookmarksnumbered=true,
            breaklinks=true,
            pdfpagemode=Fullscreen,
            pdfstartview=FitBH]{hyperref}

\usepackage{esint}

\usepackage[normalem]{ulem}
\usepackage{color}

\definecolor{Orange}{cmyk}{0,0.61,0.87,0}
\definecolor{JungleGreen}{cmyk}{0.99,0,0.52,0}
\definecolor{OliveGreen}{cmyk}{0.64,0,0.95,0.40}
\definecolor{Brown}{cmyk}{0,0.81,1,0.60}
\definecolor{RoyalBlue}{cmyk}{0.71,0.53,0,0.12}
\definecolor{Gray}{cmyk}{0,0,0,0.40}
\definecolor{LightPink}{cmyk}{0.0,0.25,0,0}
\definecolor{LLightPink}{cmyk}{0.0,0.10,0,0}
\definecolor{LightBlue}{cmyk}{0.25,0,0,0}
\definecolor{LightGray}{cmyk}{0,0,0,0.2}

\usepackage{xcolor}
\definecolor{gesfpurple}{rgb}{0.47,0.19,0.42}

\definecolor{gesflanse}{rgb}{0.00,0.50,0.50}

\definecolor{gesfblue}{rgb}{0.08,0.42,0.76}

\definecolor{gesfred}{rgb}{1,0,0}

\definecolor{gesfwhite}{rgb}{1,1,1}

\definecolor{gesfblack}{rgb}{0,0,0}

\newcommand{\geqn}[1]{Eq.\,\hypersetup{linkcolor=blue}(\ref{#1})\hypersetup{linkcolor=blue}}
\newcommand{\gfig}[1]{{\hypersetup{linkcolor=violet}Fig.\,\ref{#1}\hypersetup{linkcolor=blue}}}

\graphicspath{{figs/}}

\begin{document}

\title{Superconducting Cloud Chamber}

\author{Bo Gao}
\email{gaobo\underline{\,\,\,}79@sjtu.edu.cn}
\affiliation{Tsung-Dao Lee Institute,  Shanghai Jiao Tong University, 201210, China}

\author{Jie Sheng}
\email[Corresponding Author: ]{jie.sheng@ipmu.jp}
\affiliation{
Kavli IPMU (WPI), UTIAS, University of Tokyo, Kashiwa, 277-8583, Japan}
\affiliation{Tsung-Dao Lee Institute,  Shanghai Jiao Tong University, 201210, China}

\author{Tsutomu T. Yanagida}
\email{tsutomu.tyanagida@gmail.com}
\affiliation{
Kavli IPMU (WPI), UTIAS, University of Tokyo, Kashiwa, 277-8583, Japan}
\affiliation{Tsung-Dao Lee Institute,  Shanghai Jiao Tong University, 201210, China}

\begin{abstract}

We propose a new particle-trajectory detector composed of Josephson junctions, named the superconducting cloud chamber. By measuring the quantum phase difference, this device can detect charged particles with extremely low kinetic energy, providing a new method for detecting slow-moving particles. It can also be utilized to detect millicharged dark matter particles thermalized with the Earth's environment within the mass range of $10^3\sim 10^{10}\,$GeV.

\end{abstract}

\maketitle 

{\bf Introduction} -- 
Millicharged particles can be an interesting dark matter (DM) candidate~\cite{Holdom:1985ag,Feldman:2007wj,McDermott:2010pa,Izaguirre:2015eya,Agrawal:2016quu} because millicharge naturally ensures the stability of particle. 
As long as the millicharge of DM is large enough, it can reach thermal equilibrium with the Earth environment through long-range electromagnetic interactions with standard model (SM) particles~\cite{Neufeld:2018slx,Pospelov:2020ktu,Leane:2022hkk,Berlin:2023zpn}. Consequently, the kinetic energy of DM becomes as tiny as the Earth's temperature ($\sim 0.02\,$eV), which prevents it from surpassing the energy threshold of conventional DM direct detection experiments and thus evading experimental constraints~\cite{Pospelov:2020ktu,Berlin:2023zpn}.
Additionally, these strongly-interacting millicharged DM particles can be thermally produced in the early Universe with
a small abundance~\cite{Aboubrahim:2021ohe}. The cosmological observations are not sensitive to the DM with a small fraction~\cite{Dubovsky:2003yn,dePutter:2018xte,Kovetz:2018zan,Buen-Abad:2021mvc}.

Although many new experiments are proposed to detect millicharged DM~\cite{Pospelov:2020ktu,Kim:2007zzs,Moore:2014yba,Afek:2020lek,Budker:2021quh}, there are still a lot of parameter spaces with large mass and small fraction for the millicharged DM that need exploration~\cite{Budker:2021quh}. Moreover, current particle detectors are unable to directly observe extremely low-velocity particles, even those SM charged particles, without acceleration.
In this paper, we design a novel superconducting cloud chamber made of RF SQUIDs (radio frequency superconducting quantum interference devices~\cite{34120}) to target low-speed charged particles. 
We use natural units with $c = \hbar = 1$ throughout the text.

{\bf I. Phase Difference Induced from Charged Particles} --
A Josephson junction (JJ) consists of two superconductors separated by a thin insulating barrier.
Below the critical temperature, electrons form 
Cooper pairs and condense into a coherent state. 
These two superconducting states on the different sides of the insulator, $\ket{1}$ and $\ket{2}$, can be described by macroscopic wave functions, $\Psi_{1,2} = \sqrt{n_{1,2}} e^{i \phi_{1,2}}$~\cite{Landau:1950lwq}. 
A tunneling current occurs
when there is a phase difference between the two states, 
$\Delta \phi \equiv \phi_1 - \phi_2 \neq 0$. 
This is the so-called \textit{Josephson effect}~\cite{Josephson:1962zz,Josephson:1974uf}. 
The Josephson current exhibits periodic dependence on the phase,
$I = I_c \sin (\Delta \phi)$ where $I_c$ is the constant critical current.

\begin{figure}[!t]
\centering
 \includegraphics[width=0.47
 \textwidth]{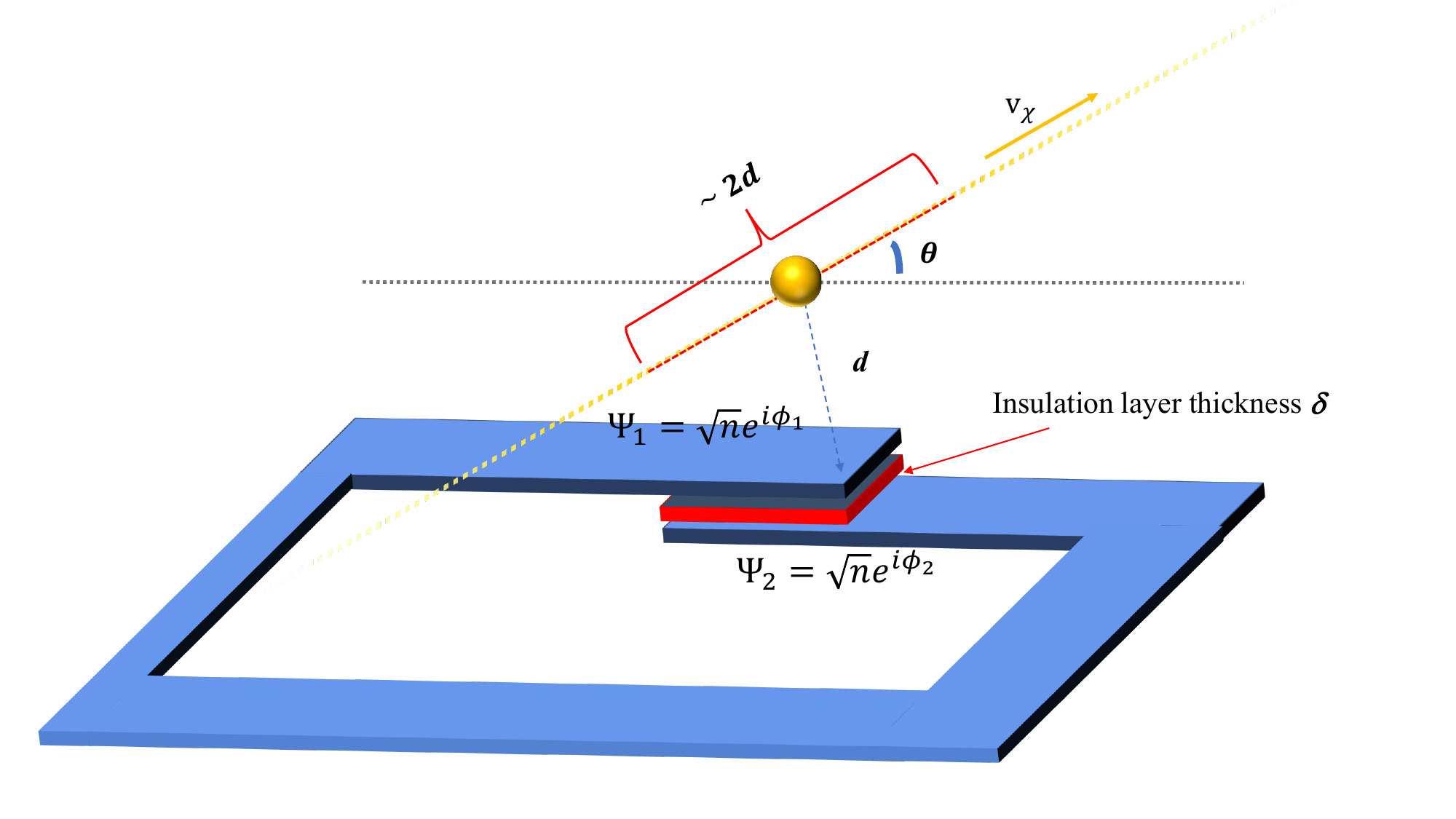}
\caption{
Schematic diagram illustrating the detection of moving charged particles by an RF-SQUID. 
The particle trajectory is at a distance $d$ from the insulator with angle $\theta$.}
\label{fig:singleUnit}
\end{figure}

If the quantum states on each side of the insulator feel a voltage difference\footnote{Voltage can also be replaced by any interaction potential involving electrons, such as gravitational potential~\cite{Jain:1987zz,Ummarino:2020loo,Christodoulou:2023hyt,Cheng:2024zde} and the fifth-force potential~\cite{Cheng:2024yrn}.} $\Delta V$, their phase difference can evolve over time as $d\phi/dt = e^* \Delta V$ where $e^* = 2e$ is the Cooper pair charge. A charged particle passing by a JJ can induce a potential and phase difference across the junction.
As shown in \gfig{fig:singleUnit}, an unknown charged particle $\chi$ flies past the junction at a distance of $d$ with a velocity $v_\chi$. 
For generality, let the particle trajectory form an angle 
$\theta$ with the JJ plane.
Assuming the particle carries a charge $\epsilon e$, there exists a distance-dependent potential around it, with a magnitude of $V(r) = \epsilon e/(4 \pi r)$. Due to the thickness $\delta$ of the insulator in JJ, there is a distance difference of $\delta$ between the superconductors on either side of the insulating layer and the charged particle.
Since the electric field is shielded inside
superconductors, a voltage difference only exists across the junction. 
This voltage is also a function of time because of the motion of charged particle. 
When the distance $d$ is much larger than the planar 
dimensions of the JJ, the accumulated phase difference is the integration over the entire time domain as,  
\begin{equation}
    \Delta \phi 
=
    \frac{\epsilon e^2}{2\pi}
    \int \frac{1}{\sqrt{d^2 + v_\chi^2 t^2}} - \frac{1}{\sqrt{(d+\delta \cos \theta)^2 + v_\chi^2 t^2}} dt.
\label{signalint}
\end{equation}
When the trajectory is parallel to the plane, the asymmetry between the two superconductors is maximal, giving the largest signal. For $\theta = \pi/2$, the superconductors are symmetric to the trajectory and the phase evolution during the entry and exit of particle cancels.

The thickness of the insulating layer is typically $\delta=1$\,nm, which is tiny compared to the overall size of a JJ ($\sim 10\,\mu$m) and the distance $d$ (The distance $d$ is constrained by geometrical aspects of the set-up around $d \simeq 1\,$mm as will explained in Sec. IV.
With $\delta \ll d$, the above integration can be calculated analytically,
\begin{equation}
    \Delta \phi = \frac{\epsilon e^2}{\pi }  \frac{\delta \cos \theta}{v_\chi d }.
\label{deltaphi}
\end{equation}
Although the integration in \geqn{signalint} is performed from $t \rightarrow -\infty$ to $t \rightarrow + \infty$, the main contribution to the integral comes from the region where the particle approaches the JJ with its horizontal distance from the JJ  is smaller than $d$ (the red dashed line in \gfig{fig:singleUnit}). During this distance, the travel time of $\chi$ is $\tau \simeq d/v_\chi$. Consequently, the flying charged particle induces a pulsed signal with frequency $f_{\text{signal}} = 1/\tau \simeq v_\chi / d$.

When a phase difference in an RF SQUID results in a Josephson current, due to geometrical self-inductance $L_S$, a magnetic flux signal $\Delta \Phi$ will also be generated within the loop.
The relationship between phase difference and the perturbative magnetic flux is, 
$\Delta \phi = 2e \oint \vec{A} d\vec{s} = 2e \Delta \Phi$, and thus\footnote{Notice that the magnetic field of the low-speed particle is a secondary effect compared to its electric field, $|{\bf B}|/|{\bf E}| = v < 10^{-6}$ for our targets. The magnetic flux here is actually induced by the electric field of passing charged particles and the self-inductance of RF SQUID.}~\cite{gross2016applied}, 
\begin{equation}
    \frac{2\pi \Delta \Phi}{\Phi_0} \simeq \Delta \phi \quad \rightarrow \quad \Delta \Phi \simeq \frac{\Phi_0 \Delta \phi}{2\pi}.
\label{eq:solu1}
\end{equation}

{\bf II. RF SQUID as Particle Detector} -- 
To detect the randomly occurring phase difference, we utilize the fact that a JJ behaves as a nonlinear inductance. The Josephson inductance, $L_J = L_c \sec(\phi)$, is a function of the phase difference $\phi = \phi_0 + \Delta \phi$ across the junction, where $L_c = \Phi_0/(2 \pi I_c)$ and $\Phi_0 \equiv  \pi/ e$. 
Due to the perturbation from passing charged particles, a small phase difference $\Delta \phi$ develops across the junction, inducing a change in the Josephson inductance. 
The change in inductance can be detected by embedding the JJ into a superconducting loop with loop inductance $L_S$, 
forming an RF-SQUID (the blue circuit in \gfig{fig:singleUnit}). We choose $\lambda \equiv L_S/L_C < 1$ to ensure a non-hysteretic RF-SQUID.

\begin{figure}[!t]
\centering
 \includegraphics[width=0.47
 \textwidth]{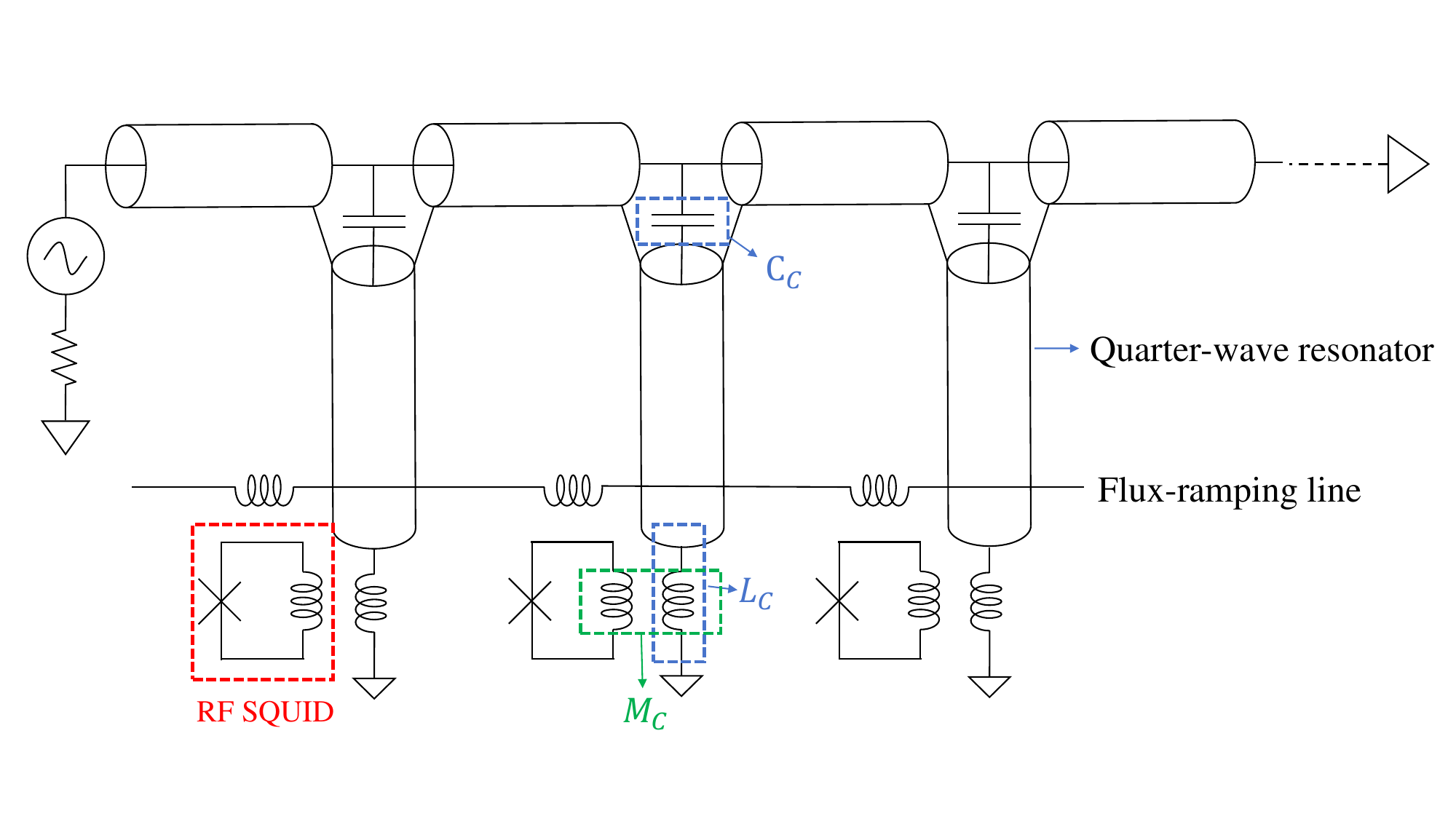}
\caption{A schematic representation of a multiplexed readout module for the superconducting cloud chamber. The module comprises multiple microwave resonators capacitively coupled to a common microwave feedline. 
Each resonator is modulated by its own RF-SQUID, which serves as the fundamental particle detection unit. 
The RF-SQUID is inductively coupled to the load impedance of the quarter-wave resonator. The microwave signal transmitted through the feedline is amplified by a HEMT amplifier. A common flux-ramp modulation line is applied to all RF-SQUIDs to linearize their responses. }
\label{fig:wires}
\end{figure}

The general configuration and detectability of the chamber can be realized by a proposed microwave readout schema shown in \gfig{fig:wires}. 
It was originally developed for large transition edge sensor (TES) arrays designed for cosmic microwave background observation and high-resolution X-ray/Gamma-ray spectrometers, where an RF-SQUID coupled to a TES pixel acts as a nonlinear flux-actuated inductor~\cite{2021ApPhL.118f2601D,6994798,256711,2017ApPhL.111f2601M,HOLMES:2019ykt,lens.org/027-935-516-419-894,SathyanarayanaRao:2020itw}.  
The RF-SQUID is coupled to the load inductor of a quarter-wave resonator, such that the input signal from the TES modulates the resonance frequency of resonator. In our proposal, charged particles directly interact with the JJ in the RF-SQUID, inducing a phase variation $\Delta \phi$ across the junction and altering its inductance. 
This results in a measurable resonance frequency shift of the quarter-wave resonators from $f_0$ to $f(\Delta \phi)$. In the small signal limit, the shift is~\cite{Wegner:2021yrn}, 
\begin{equation}
    f(\Delta \phi) = f_0 - 4 f_0^2 \left(C_C Z_0 + \frac{L_C}{Z_0} - \frac{M_C^2}{L_S Z_0} \frac{\lambda \cos \Delta \phi}{1+\cos \Delta \phi} \right).
\end{equation}
Here, $f_0$ is the first mode resonance frequency of the unloaded resonator, $C_C$ is the capacitance of microwave resonators, $Z_0$ is its characteristic impedance, 
and $M_C$ is the mutual inductance between RF-SQUID and quarter-wave resonator with 
load impedance $L_C$.
By measuring the resonance frequency shift, the phase variation $\Delta \phi$ induced by the charged particles can be inferred. 

In practice, a specialized readout technique known as the flux ramp has been developed, which applies a sawtooth waveform to the RF-SQUID via a flux-ramping line. The ramp frequency of the sawtooth waveform, $f_{\text{ramp}}$, is chosen to be at least twice the frequency of the input signal. The amplitude of the waveform is set such that it induces an integer number of flux quanta in the RF-SQUID. This waveform periodically modulates the response of the RF-SQUID at a modulation frequency $f_{\text{mod}} =n f_{\text{ramp}}$, and consequently modulates the resonance frequency of the quarter-wave resonator. Since $f_{\text{mod}}$ is much higher than the signal frequency, the influence of the input signal appears as a phase shift in the modulated SQUID response. The signal of a charged particle can be extracted from this phase shift. Further details are provided in Appendix A.

{\bf III. Sensitivity} -- 
The resolution of the phase variation $\Delta \phi$ induced by the passing charged particles is determined by the flux noise of the readout module.
Since the RF-SQUID used here is dissipationless and there is no input filter, 
the primary noise contributions are from:
(1) two-level system (TLS) noise of the quarter-wave resonator, (2) Johnson noise of the microwave amplifier, (3) $1/f$ noise of the RF-SQUID, and (4) other energy loss factors in the readout circuit \cite{Kohjiro:2014wn,Schuster:2022grf,Yu:2022dso,GarciaRedondo:2024fjk}. 
State-of-the-art readout modules have achieved a system white noise of less than
$1 \mu \Phi_0/\rm{Hz}^{1/2}$~\cite{2020ApPhL.117l2601N,Malnou:2023wfo,Neidig:2025tfs}. Additionally, low-frequency noise is further suppressed by the flux ramp modulation technique \cite{2017SuScT..30f5002K,Schuster:2022grf}.

As discussed above, the signal frequency is bounded by the ramp frequency $f_{\text{ramp}}$, limited by the resonator bandwidth. We take $f_{\text{ramp}} = 1\,$MHz as a benchmark. At low frequencies, $1/f$ noise dominates below the corner frequency $f_{\text{knot}} \simeq 10\,$Hz. Thus, the detectable signal band is $(10, 10^6)\,$Hz, within which the readout chain is band-limited and higher-frequency noise is discarded. The total readout noise in this range can be estimated as
\begin{equation}
    \Phi_{\text{n}} \equiv 1 \mu \Phi_0/\sqrt{\text{Hz}} \times \sqrt{f_{\text{ramp}}} = 10^{-3}\times \Phi_0. 
    \label{eq:sen}
\end{equation}
The estimated noise floor is consistent with the system noise levels observed in state-of-the-art microwave multiplexed readout systems currently used for TES array readout
\cite{2020ApPhL.117l2601N,Malnou:2023wfo,Neidig:2025tfs}.

Based on the frequency relationship of the signal, $f_{\text{signal}} = v_\chi/d$, the aforementioned signal-frequency-range can be further converted into the detectable velocity range of the particles, $ d/10^{-1}\,\text{s} < v_\chi < d/10^{-6}\,\text{s}$. In our setup, the spacing $d = 1\,$mm and thus the corresponding detectable particle velocity is quite small as, $v_\chi \subset (3 \times 10^{-11}, 3 \times 10^{-6}) = (0.01\,\text{m}/\text{s}, 1000\,\text{m}/\text{s})$.

Since the noise $\Phi_n$ in \geqn{eq:sen} is expressed in terms of magnetic flux, the detectability of a signal should also be assessed by comparing it to the magnetic flux signal $\Delta \Phi$ generated by the charged particle in \geqn{eq:solu1}. 
With $f_{\text{ramp}} = 1\,$MHz, the signal-to-noise ratio (SNR) can therefore be estimated as,
\begin{equation}
    \text{SNR} \equiv \frac{\Delta \Phi}
    {\Phi_n}
    =
    \frac{\Delta \Phi}
    {10^{-3} \times \Phi_0}.
\label{SNR}
\end{equation}
Conservatively, we consider the criterion for detectable signal is that $\text{SNR} \geq 10$~\cite{Qin:2023urf}.

\begin{figure}[!t]
\centering
 \includegraphics[width=0.47
 \textwidth]{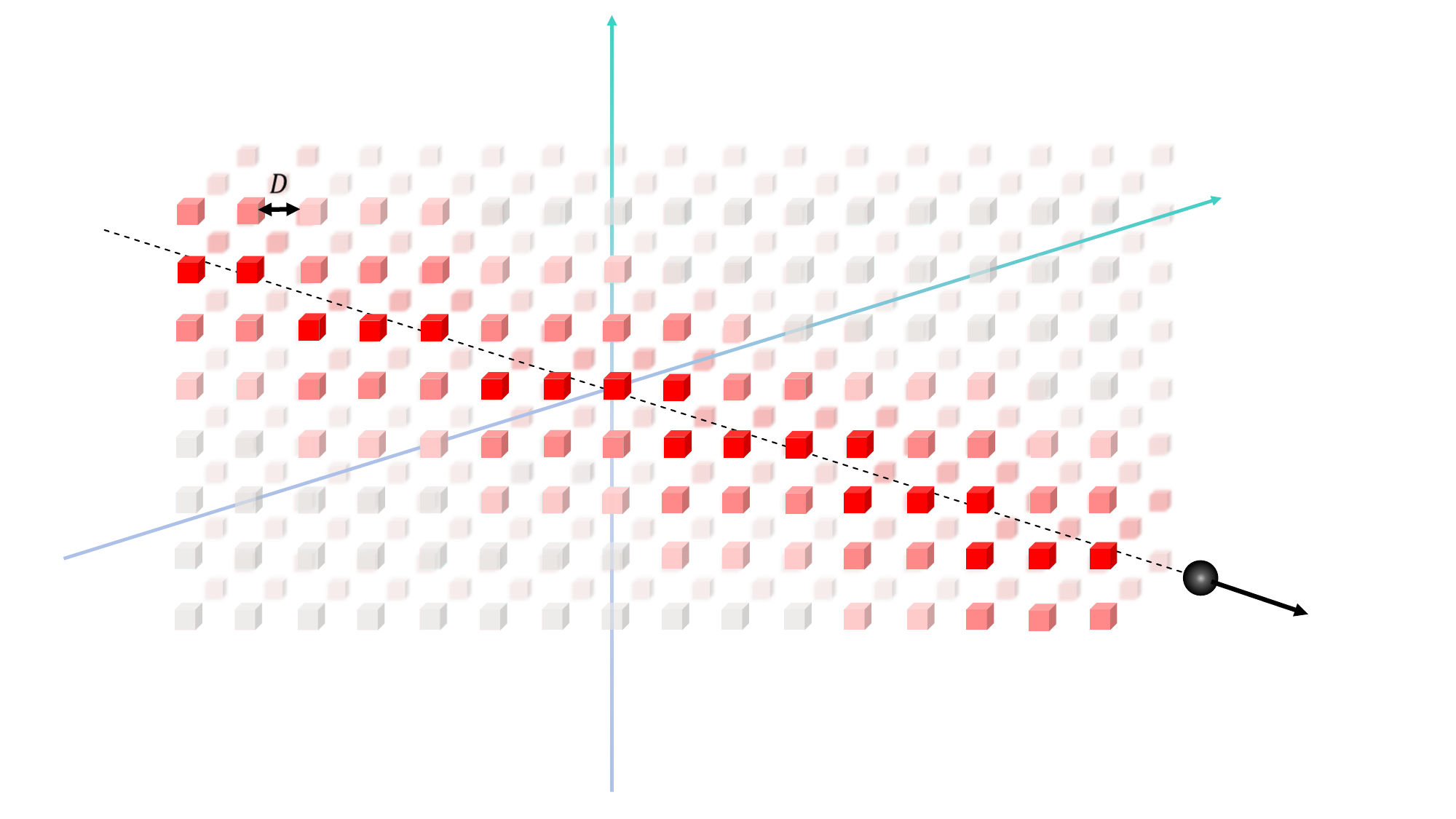}
\caption{
The schematic diagram of the superconducting cloud chamber. Each cube represents a basic detection unit shown in Fig.1, forming a 3D array with a spacing of $D$ between adjacent units. When a charged particle passes through the cloud chamber, it excites pulsed signals in JJs surrounding its trajectory. 
}
\label{fig:chamber}
\end{figure}

{\bf IV. Superconducting Cloud Chamber} -- Based on the particle-detector unit made of an RF SQUID and the readout schema in \gfig{fig:wires}, we propose a superconducting cloud chamber composed of a 3D array of these units for visualizing the trajectories of charged particles. 
As shown in \gfig{fig:chamber}, each cube represents an RF SQUID shown in \gfig{fig:singleUnit}, spaced by $D$ in all three directions. 
The geometrical size of a non-hysteresis RF-SQUID must be small, on the order of $\mathcal{O}(10)\,\mu$m, and the dimension of JJ is $a \simeq 3\,\mu$m. We can design the spacing $D$ in the array to be tens of times larger than the size of a single unit, such as $D \simeq 3\,$mm. This means each JJ is surrounded by an exclusion volume $(3\,\text{mm})^3$, making the probability of a particle passing in close proximity to a JJ extremely low. For instance, the probability for a DM particle to approach a JJ within a distance $10 a$ is $(30\,\mu\text{m})^3/(3\,\text{mm})^3 = 10^{-6}$. 
In this way, if a particle passes through the cloud chamber, its distance $d$ from the closest SQUIDs will be fixed roughly as $d \sim 1\,$mm $\lesssim D$, validating the approximation used in \geqn{signalint}.

For a cube-shaped cloud chamber with a volume of ($3\,$cm)$^3$,  
$10 \times 10 \times 10 = 10^3$ SQUIDs are required in total. The maximum number of SQUIDs in a readout module is discussed in Appendix B.
The design of the cloud chamber enables clear trajectory reconstruction.
The detectors closest to the particle will measure the strongest signal. The remaining detectors will register progressively weaker signals as the distance increases. 
As depicted in \gfig{fig:chamber}, the black dashed line is the trajectory of a charged particle. 
The varying shades of red represent the strength of the signal, with darker red indicating a stronger signal.

Furthermore, as previously mentioned, the pulsed signal is only generated when the particle are close to the SQUID. Along the trajectory of the particle, the signals from the detector occur almost sequentially in time.
Therefore, the velocity of the particle can also be reconstructed by analyzing the signal strength and time sequence of detectors.

\begin{figure}[!t]
\centering
 \includegraphics[width=0.47
 \textwidth]{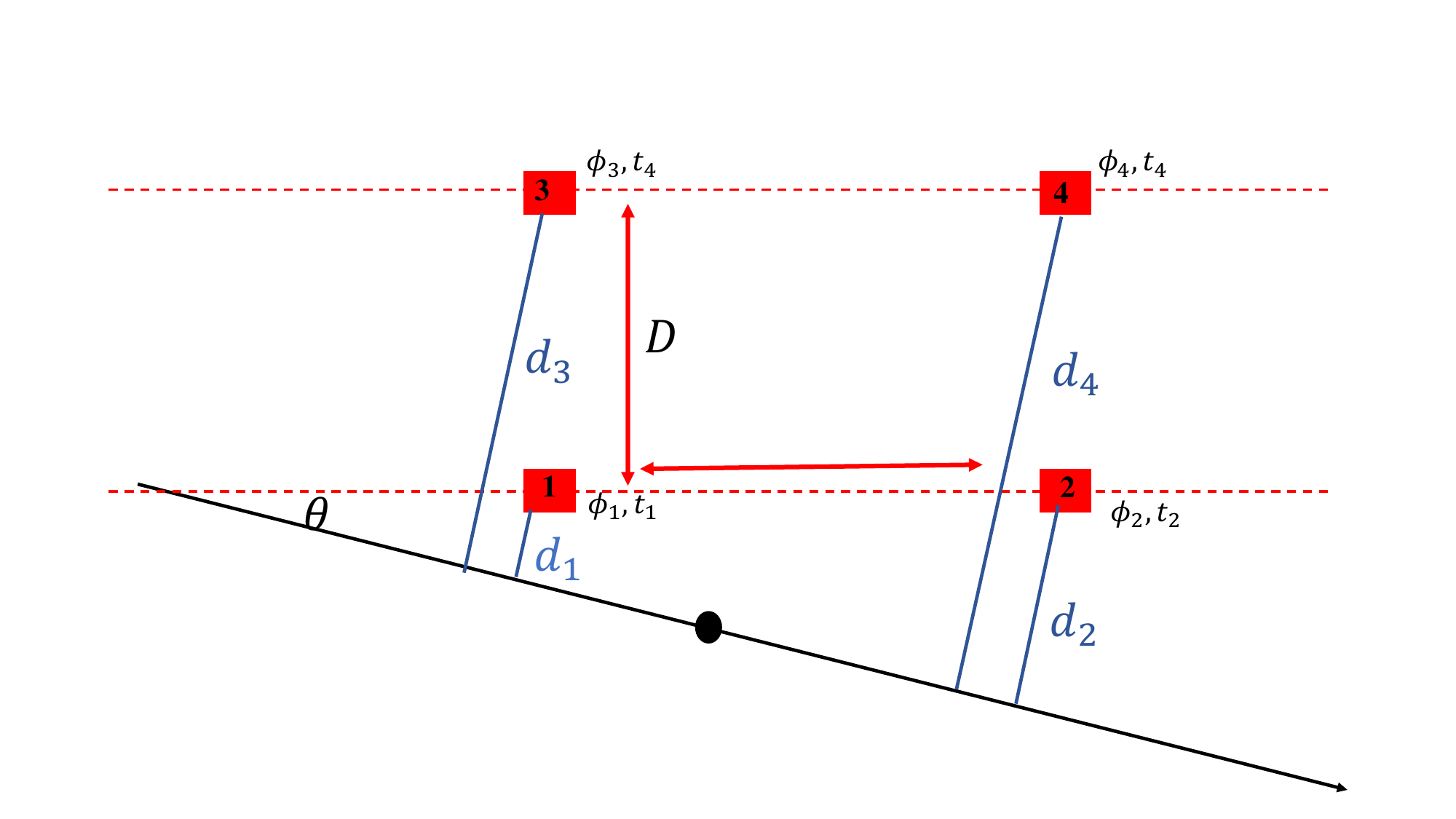}
\caption{A schematic illustration of a particle passing through the superconducting cloud chamber (local view).}
\label{fig:ana}
\end{figure}

For example, as shown in \gfig{fig:ana}, the charged particle flies by the detector $1 \sim 4$ with distance 
$d_i (i = 1 \sim 4)$ and flying angle $\theta$ as defined above.
Naturally, the relationship between these distances and angle can be established,
\begin{equation}
    d_3 = d_1 + \frac{D}{\cos \theta} \,, \quad 
    d_4 = d_2 + \frac{D}{\cos \theta}.
\label{angle_dis}
\end{equation}

The detectors would see the phase difference $\phi_i (i = 1 \sim 4)$ 
as $\phi_i = \epsilon e^2 \delta \cos \theta/ \pi v_\chi d_i$.
The signal peak arises at the moment $t_i (i = 1\sim 4)$. 
One can also establish the relationship between the signal timing and the particle velocity,
\begin{equation}
    t_2 - t_1 = \frac{D}{\cos \theta v_\chi}.
\label{time}
\end{equation}
To see the time difference, the timing resolution between different detectors must be better than, $\delta t < t_2 - t_1 \simeq D/v_\chi$. Since $D \sim 1\,$mm and $v_\chi \subset (0.01, 1000)\,$m/s, the time resolution should be better than $\delta t < 10^{-5}\,$s.
This condition is easily achieved by the current timing technology.

Including the four phase equations together with \geqn{angle_dis} and \geqn{time}, we can list a total of seven equations. 
These equations resolve the degeneracies in the signal, allowing us to determine the the particle charge $\epsilon$, velocity $v_\chi$, angle $\theta$, and the distances $d_i$.
Thus, besides detecting the existence, this chamber also determines the particle nature. The precision depends on the signal-to-noise ratio, for
$\text{SNR}=10$, the uncertainties of a single detector are below $10\%$. Performing a combined analysis with multiple detectors can further reduce the errors.

The superconducting cloud chamber can also effectively eliminate the majority of environmental backgrounds. 
Firstly, random system backgrounds only sporadically generate signals and do not present particle trajectories. Secondly, since the detector is sensitive only to low-speed flying charged particles, 
signals generated by high-energy SM particles like cosmic rays cannot be detected, thus not forming background noise\footnote{It is preferable to place the detector in an underground laboratory to shield most of the flying particles. Although cosmic muons may be decelerated by Earth shielding, they quickly decay to electrons and the speed of electrons is too high to leave signal in chamber setup.}. 
Lastly, once clear particle trajectories are identified in the chamber, confirming the presence of flying particles, we can also determine the charge of particle by analyzing the signal intensities. In the case of millicharged DM detection, the charges of SM charged particles in the background are all integers, making them easily distinguishable. A detailed discussion on determining the velocity and charge of particles using the chamber can be found in Appendix.C.

\begin{figure}[!t]
\centering
 \includegraphics[width=0.47
 \textwidth]{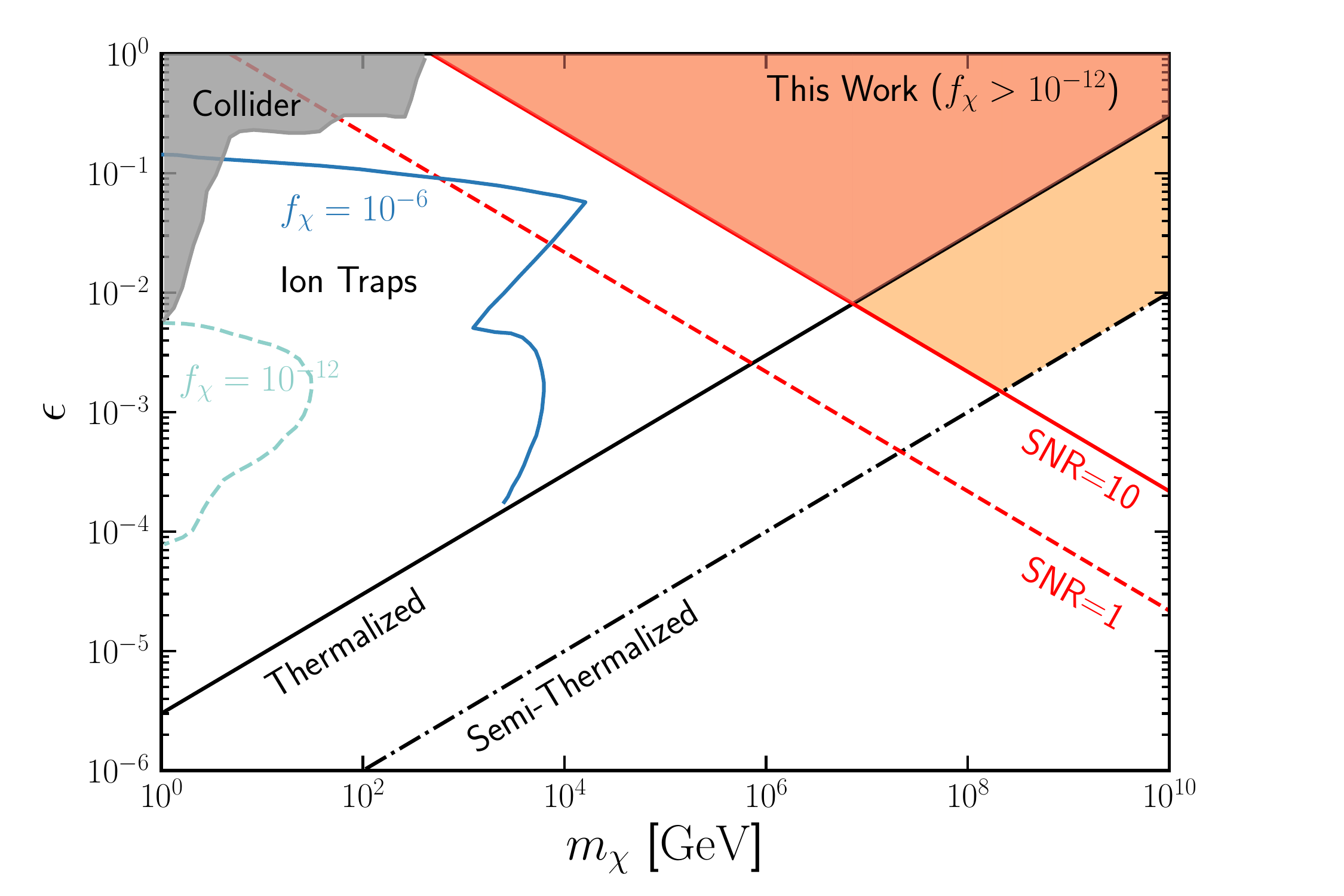}
\caption{The projected sensitivity of superconducting cloud chamber on millicharged DM assuming the $\text{SNR} = 10$, shown as the different colored shaded region depending on if the DM is thermalized or not. The constraint from collider is the grey shaded region, while the constraints from ion trap experiments are blue solid and green dashed lines depending on the DM fraction. The sensitivity by assuming $\text{SNR}=1$ is also presented as the red dashed line.}
\label{fig:limit}
\end{figure}

{\bf V. Application to Millicharged DM} -- One application of this chamber is the detection of millicharged DM, which is assumed to directly couple only to massless SM photons.
Previous research~\cite{Neufeld:2018slx,Pospelov:2020ktu,Leane:2022hkk,Berlin:2023zpn} has shown that millicharged DM can reach thermal equilibrium through multiple scatterings with the Earth's environment at depths of $1\,$km underground (the black solid line in \gfig{fig:limit}), once its electric charge $\epsilon$ satisfies the condition~\cite{Pospelov:2020ktu},
$\epsilon \gtrsim 3\times 10^{-6} \sqrt{m_\chi / \rm{GeV}}$. As a result, the kinetic energy of DM equals the Earth's temperature, $T_\chi \simeq 0.02\,$eV. The velocity of DM relates to its mass as $v_\chi = \sqrt{2 T_\chi/ m_\chi}$.

The detectable velocity range of a charged particle in the superconducting cloud chamber,
$(3 \times 10^{-11}, 3 \times 10^{-6})$, corresponds to a detectable range of the millicharged DM masses, $m_\chi \sim (1,10^{10})\,$GeV.
Assuming the SNR $= 10$, the minimum magnetic flux that the detector can measure is, $\Delta \Phi \geq 10^{-2} \times \Phi_0 $, according to \geqn{SNR}. Compare it with the signal \geqn{eq:solu1}, one can get the sensitivity to the phase difference $\Delta \phi \geq \pi / 50$ and then the millicharge, 
\begin{equation}
    \epsilon \geq 3.4
    \times 10^6 \times v_\chi
    = 
    21.8 \times \left(\frac{\rm{GeV}}{m_\chi}\right)^{1/2}
    ,
\end{equation}
which is the region above the red solid line in \gfig{fig:limit}. 
Here, we take the $\cos\theta$ to be its statistical average $2/\pi$.
Since this limit only applies to the DM thermalized with the Earth environment, its overlap (red shaded region) with region above black line is the sensitivity of the present superconducing cloud chamber to millicharged DM.

Furthermore, because of the long-time (the age of the Earth) accumulation effects, the number density of the millicharged DM in the earth also significantly increases. 
Depending on the fraction of the millicharged DM in the total DM $f_\chi$, 
its number density is,
$ n_\chi = 3 \times 10^{15}\,$cm$^{-3}\times f_\chi (1\,\rm{GeV}/m_\chi)$~\cite{Pospelov:2020ktu}. To ensure the detection rate to be larger than $1\cdot\,$day$^{-1}$ in $1\,$cm$^3$ chamber, the DM density $n_\chi$ should be, 
$ n_\chi \gtrsim 10^{-7}/{\rm cm}^3 \times \left( 10^{-8} /v_\chi \right).
$
Even for the heaviest DM mass $m_\chi = 10^{10}\,$GeV, the superconducting cloud chamber has a potential to detect the millicharged DM with a tiny fraction of $f_\chi > 10^{-12}$.

For the DM that did not fully reached thermalization at a distance of $1\,$km inside the Earth (black dashed line), a small fraction of them still have lower velocities due to distribution effects. Since the chamber has very low requirements for the DM number density, this parameter space (orange shaded region) might also be detectable.

Due to its extremely low kinetic energy, the millicharged DM thermalized in Earth cannot surpass the threshold for traditional DM direct detections~\cite{Budker:2021quh,ArguellesDelgado:2021lek,PandaX:2023toi,Iles:2024zka}, necessitating novel experimental methods for its detection, such as electrostatic accelerator~\cite{Pospelov:2020ktu}, oil-drop experiment~\cite{Kim:2007zzs}, levitation experiments with microspheres~\cite{Moore:2014yba,Afek:2020lek}, ion traps~\cite{Budker:2021quh}, and so on~\cite{SuperCDMS:2020hcc,Chen:2022abz,CUORE:2024rbd,Ema:2024oce}. For a DM fraction of $f_\chi = 10^{-6}$ and $f_\chi = 10^{-12}$, the most stringent constraints from ion-trap experiments are shown as the blue solid and green dashed lines~\cite{Budker:2021quh}. Because of the decrease in energy transfer efficiency and number density, this limit is not sensitive to high-mass and small-fraction parameter ranges. We also demonstrate collider search and beam dump experiment constraints as the gray region~\cite{1995ZPhyC..67..203A,Prinz:1998ua,Magill:2018tbb,Berlin:2018bsc,Kelly:2018brz,Harnik:2019zee,ArgoNeuT:2019ckq,Ball:2020dnx,Marocco:2020dqu}. 
It can be observed that the superconducting cloud chamber has a unique advantage in the high-mass parameter range, as higher-mass millicharged DM particles have slower velocities, aligning perfectly with the sensitivity range of the detector.

{\bf Discussions and Summary} --
Particle tracking detectors play a crucial role in the fields of particle physics. 
Many fundamental charged particles such as the positron and muon were discovered in Wilson cloud chamber~\cite{Grupen_Shwartz_2008}.
Afterwards, the bubble chamber made a series of significant discoveries~\cite{Glaser:1952zz} including weak neutral currents.
Cherenkov detectors can also be used to detect neutral high-speed particles, such as high-energy neutrinos~\cite{Bolotovskii_2009}.
Recently, new tracking detectors such as wire chambers, spark chambers, drift chambers, and solid-state nuclear track detectors have been developed for a wider range of physics explorations~\cite{Sauli:1977mt,GUO2012233}. 
DM is typically considered as an unknown particle. Its detection can also be aided by various bubble chambers~\cite{PICASSO:2012ngj,COUPP:2012jrk,PICO:2015amc,PICO:2023uff}.

The detection methods based on scattering and ionization used by the detectors above all have certain energy thresholds. Ionizing an atom typically requires $\sim 10\,$eV of energy. 
This requires the kinetic energy of the incident particles to be greater than this threshold. 
Consequently, 
they can not detect the millicharged DM thermalized with the Earth's environment and even the SM charged particles with velocity lower than $10^{-4}$.

In this paper, we propose a cloud chamber made of JJs. Due to the absence of resistance in superconductors, the energy threshold required to excite superconducting current signals is extremely low.
The proposed setup may face challenges in array construction and timing synchronization. Since this idea is entirely new, these aspects have not yet been attempted, but they are technically feasible in principle.
Considering the background of circuit readout, the superconducting cloud chamber can detect charged or millicharged particles traveling at speeds of $v_\chi \sim (10^{-11}, 10^{-6})$. 
Even if it constitutes only $f_\chi \gtrsim 10^{-12}$ of DM, the present superconducting cloud chamber is sensitive to millicarged DM with masses of $(10^3, 10^{10})\,$GeV. Furthermore, this chamber can also be used to track the movement trajectories of charged particles in space or specific laboratories, opening up new avenues for the detection of low-speed charged 
 particles.

\section*{Acknowledgements}

The authors thank Tie-Sheng Yang for improving the figures. They also thank Hong Ding, Shao-Feng Ge, Shigeki Matsumoto, and Ning Zhou for fruitful discussions. 
B. G. is supported by the Innovation Program for Quantum Science and Technology (No.2021ZD0302700).
J.S. is supported by the Japan Society for the Promotion of Science (JSPS) as a
part of the JSPS Postdoctoral Program (Standard) with grant number: P25018, and by the World
Premier International Research Center Initiative (WPI), MEXT, Japan (Kavli IPMU).
T. T. Y. is supported by the Natural Science Foundation of China (NSFC)
under Grant No. 12175134, MEXT KAKENHI Grants No. 24H02244, and World Premier International Research Center Initiative
(WPI Initiative), MEXT, Japan.

\providecommand{\href}[2]{#2}\begingroup\raggedright\endgroup

\begin{appendix}
\section{Appendix A -- Detection of the Phase Variation}
\label{appA}

The RF-SQUID exhibits a non-linear response to charged particles. 
To linearize this response, a flux-ramp modulation technique is employed. A sawtooth signal with frequency 
$f_{\rm ramp}$ and amplitude $n \Phi_0$ (where n is an integer) is applied to the common flux-ramping line. This induces periodic oscillations in the SQUID response. 
If the slew rate of the applied ramp exceeds that of any input signal, variations in the input signal appear as phase offsets during the ramp period, producing a phase shift in the SQUID response relative to its free oscillation state (i.e., without the interaction of charged particles). This phase shift can be calculated from a Fourier measurement of the SQUID response,
\begin{equation}
    \phi = \arctan\left(
    \frac{-\sum \theta(t) \sin (2 \pi f_{\text{mod}} t)}
    {-\sum \theta(t) \cos (2 \pi f_{\text{mod}} t)}
    \right).
\end{equation}
Here, $\theta(t)$ represents the discrete sampling of the SQUID response, and $f_{\text{mod}} = n \times f_{\rm ramp}$ is the modulation frequency. Typically, $f_{\text{mod}}$ is required to be at least $10$ times the maximum signal frequency.

\section{Appendix B -- Multiplexing Factor}

The number of RF SQUIDs in a readout module, which we define as the multiplexing factor $M$, 
can be estimated using as follows. Let $S_f$ be the frequency spacing between resonator tones, and 
$f_{\rm ADC}$ be the ADC bandwidth, the multiplexing factor is then defined as $M \equiv f_{\rm ADC}/S_f$. Several constraints apply to $S_f$~\cite{HOLMES:2019ykt}: (1) $S_f$ must be significantly larger than the resonator
bandwidth, $S_f > g_f \times \Delta f_{\rm BW}$, where $\Delta f_{\rm BW}$ is the resonantor bandwidth and $g_f$ is a guard factor; (2) The Nyquist-Sannon sampling theorem requires $\Delta f_{\rm BW} > 2 f_{\text{mod}}$; (3) The ramp frequency must exceed the maximum signal frequency, 
$f_{\rm ramp} > R_d/\tau$, where $\tau$ is the rise time of the charged particle (DM) signal and 
$R_d$ is a signal distortion factor. In conclusion, the multiplexing factor per ADC board is, 
\begin{equation}
    M = \frac{f_{\rm ADC} \cdot \tau}{2 n \cdot g_f \cdot R_d}.
\end{equation}

Assuming the typical values of $f_{\rm ADC} = 1\,$GHz, $n= 5$, $g_f = 5$, and $R_d = 5$, if a 
signal rise time is $10\,\mu$s, the multiplexing factor per ADC board is estimated to be $M = 40$.
If the frequency of signal becomes lower and the rise time is extended to be $1\,$ms, the multiplexing factor increases to be $M = 4000$. This multiplexing capability, which depends on the frequency of signals, has been demonstrated in TES-based CMB detector arrays~\cite{Henderson:2018jlc,Jones:2024yss}.

\end{appendix}

\vspace{15mm}
\end{document}